\documentclass[fleqn,twoside]{article}
\usepackage{espcrc2}

\usepackage{graphicx}
\usepackage[figuresright]{rotating}

\newcommand{\AmS}{{\protect\the\textfont2
  A\kern-.1667em\lower.5ex\hbox{M}\kern-.125emS}}
\newcommand{\ba}{\begin{eqnarray}}
\newcommand{\ea}{\end{eqnarray}}
\newcommand{\be}{\begin{equation}}
\newcommand{\ee}{\end{equation}}
\newcommand{\beq}{\begin{equation}} 
\newcommand{\eeq}{\end{equation}}   
\newcommand{\bea}{\begin{eqnarray}} 
\newcommand{\eea}{\end{eqnarray}}

\title{PHOKHARA 7.0 Monte Carlo generator: 
       the narrow resonances implementation and new pion and kaon form factors.\thanks{
 Supported in part by Polish Ministry of Science and High Education
   from budget for science for years 2010-2013: grant number N N202 102638 and 
the European Community-Research 
Infrastructure Integrating Activity 
``Study of Strongly Interacting Matter'' 
(acronym HadronPhysics2, Grant Agreement n. 227431)
 under the Seventh Framework Program of EU.}}

\author{Henryk Czy\.z, \address[kt]{ Institute of Physics, University of
Silesia,
PL-40007 Katowice, Poland. 
        }%
     \ \   Agnieszka Grzeli{\'n}ska\address[cr]{Institute of Nuclear Physics 
Polish Academy of Sciences, PL-31342 Cracow, Poland}
       }
       
\begin{document}

\begin{abstract}
Experiments at high luminosity 
 electron-positron colliders allow to study the kaon and the pion form
 factors in the time-like region up to high energies. Also
  the kaon and the pion pair production at and around
 the narrow resonances $J/\psi$ and  $\psi(2S)$ can be investigated,
  with the interference
 between electromagnetic and hadronic amplitudes as one of the most
 interesting phenomenas. Parameterisations of
 charged and neutral kaon as well as pion form factors,
  which lead to an improved description of the data in the region of
 large invariant masses of the meson pair, are presented.
 They are implemented into the Monte Carlo generator PHOKHARA,
  together with the
 hadronic couplings of charged and neutral kaons to  $J/\psi$ and
 $\psi(2S)$. The physics case as well as details of the implementation
 are discussed. 
\vspace{1pc}
\end{abstract}

\maketitle

\section{Introduction}

 During last years meson factories, using the radiative return method \cite{Zerwas,Binner:1999bt}, contributed a lot to the error
 reduction of the hadronic contributions to the  muon anomalous magnetic
 moment and the running electromagnetic fine structure constant
 \cite{Hoecker,Teubner}.

An important tool in these analyses is  a Monte Carlo generator
 which simulates 
all the measured reactions. To meet these requirements, the
generator EVA was developed \cite{Binner:1999bt,Czyz:2000wh}, 
which is based on leading
order matrix elements combined with structure function methods for an
improved treatment of initial state radiation. 
To improve the accuracy of these simulations the complete next-to-leading
order (NLO) QED corrections were evaluated
\cite{Rodrigo:2001jr,Kuhn:2002xg}
 and
implemented into the generator PHOKHARA
\cite{Rodrigo:2001kf,Czyz:2002np,Czyz:PH03,Nowak,Czyz:PH04,Czyz:2004nq,Czyz:2005as,Czyz:2007wi,Czyz:2008kw,Czyz:2010hj}.
 For a recent review of theoretical and experimental results
 see e.g. \cite{Actis:2009gg}.

 In the last paper, resulting in the implementation in the PHOKHARA
 generator \cite{Czyz:2010hj}, the possibility of studies of 
 the narrow resonances $J/\psi$ and $\Psi(2S)$ at B-meson factories
 was investigated in details profiting from the previous work
 on the narrow resonances \cite{Czyz:2009vj}. In this contribution 
 the results of \cite{Czyz:2009vj,Czyz:2010hj} are summarised and
 complemented with additional material not presented there. 

 The new models of pion and kaons form factors are presented in Section
\ref{formfactors}. In Section \ref{nr} the modelling of the narrow
resonances is shortly sketched. In Section \ref{phok} the implementation
 of the narrow resonances in PHOKHARA is described and a discussion
 of the importance of the 
 FSR NLO radiative corrections is presented. A short summary follows
 in Section \ref{sum}.

\section{Modeling of the pion and kaon form factors}\label{formfactors}

 For a realistic
 generation of the processes $e^+e^-\to \pi^+\pi^-\ + \ {\rm photons}$,
 $e^+e^-\to K^+K^-\ + \ {\rm photons}$ and $e^+e^-\to \bar K^0 K^0\ + \
 {\rm photons}$ models for the electromagnetic pion and kaons 
 form factors are required.
 In PHOKHARA 5.0 \cite{Czyz:2005as,Czyz:2004nq} and 6.0 \cite{Czyz:2007wi}
  models presented in \cite{Bruch}
  were implemented.
 They
 were published before
  the CLEO-~c measurement
 of the form factor in the vicinity of the $\psi(2S)$ resonance
 \cite{Pedlar:2005sj} and underestimate the experimental results
 significantly.
 Similarly the model predictions at $J/\psi$ are lower than
 the pion form factor calculated in \cite{Milana:1993wk}
  from $B(J/\psi\to\pi^+\pi^-)$ and $B(J/\psi\to e^+e^-)$ decay rates.
 In \cite{Czyz:2010hj} an effort was made to get better description
  of the form form factors, especially at high meson pairs invariant
 masses. In combination with the $J/\psi$ and $\psi(2S)$ hadronic
 amplitudes parameterisations, obtained in \cite{Czyz:2009vj} on the
 basis of new experimental data, it allows for a description of the
 pion and the kaon pairs production reflecting all existing experimental data.

  The models proposed in \cite{Czyz:2010hj} are generalisations of
 the models used in \cite{Bruch} and for both, pions and kaons, they
 are inspired by the dual QCD model of the pion form 
 factor \cite{Dominguez:2001zu}. The infinite towers of $\rho$, $\omega$
 and $\phi$ radial excitations, present in this models, are essential for
 getting the right behaviour of the form factors at high invariant
 masses of the meson pairs. 

The ansatz for the pion form factor reads
\bea
F_{\pi}(s) &=&  \left[\sum\limits_{n=0}^5 c_{\rho_n}^\pi 
BW_{\rho_n}(s)\right]_{fit} \nonumber \\
&&+  \left[\sum\limits_{n=6}^{\infty}c_{\rho_n}^\pi  BW_{\rho_n}(s)
\right]_{dQCD} \ ,
\label{piformf}
\eea
with the  parameters of the first six radial excitations fitted to the data 
and others taken
as assumed in the original model \cite{Dominguez:2001zu}.
For details we refer the reader to \cite{Czyz:2010hj}. 
In the original model the coupling constants $c_{\rho_n}^\pi$ are
real, but it is impossible to fit the data using real couplings,
 even if one allows that
 they are different from their model values. Thus they were allowed to be
 complex, keeping the normalisation (sum of all couplings $=1$)
 unchanged. With these assumptions the fit is very good
 $\chi^2/d.o.f = 271/270 $. This might be an indication that the radial $\rho$
 excitations decay to the same final states, as in this case the mixing
 between them can generate complex couplings after diagonalisation of
 the mass matrix. This subject was not studied in details as the
 experimental knowledge about the decay modes of the $\rho$ mesons is
 very poor with the exception of $\rho(770)$ \cite{Nakamura:2010zzi}.
 
The ansatz for the kaon form factors read

\bea
 F_{K^{+,0}}(s)&=&
  a_{K^{+,0}}\biggl(  \left[\sum\limits_{n=0}^{N_\rho} c_{\rho_n}^K 
BW_{\rho_n}(s)  \right]_{fit} \nonumber \\
&&\kern-20pt +  \left[\sum\limits_{n=N_\rho+1}^{\infty}c_{\rho_n}^K  BW_{\rho_n}(s)
\right]_{dQCD}
\biggr) \nonumber \\
&&+ \frac{1}{6}
 \biggl(  \left[\sum\limits_{n=0}^{N_\omega} c_{\omega_n}^K 
BW_{\omega_n}^{c}(s)\right]_{fit} \nonumber \\
&&\kern-20pt+  \left[\sum\limits_{n=N_\omega+1}^{\infty}c_{\omega_n}^K  BW^c_{\omega_n}(s)
\right]_{dQCD}
\biggr)\nonumber \\
&&+ \frac{1}{3} \biggl(  \left[\sum\limits_{n=0}^{N_\phi} c_{\phi_n}^K 
BW_{\phi_n}^{K}(s)  \right]_{fit} \nonumber \\
&&\kern-20pt+  \left[\sum\limits_{n=N_\phi+1}^{\infty}c_{\phi_n}^K  BW^{K}_{\phi_n}(s)
\right]_{dQCD}
\biggr)  \,,
\label{Kformfpm1}
\eea
 with
 \bea
   a_{K^{+}} = \frac{1}{2} \ , \ a_{K^{0}} = -\frac{1}{2} \ .
\eea
\begin{figure}[ht]
 \vspace{0.5 cm}
\begin{center}
\includegraphics[width=6.5cm,height=6.cm]{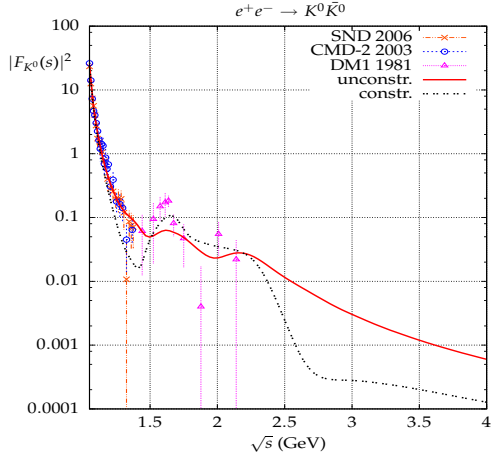}
\caption{The experimental data
 \cite{Mane:1980ep,Achasov:2001ni,Akhmetshin:2002vj,Akhmetshin:2003zn,Achasov:2006bv} 
  compared to the model fits results (see text for details).
\label{KzKz}}
\end{center}
\end{figure}

 Two versions of the model were used, one with an additional assumption
 $c_{\rho_n} = c_{\omega_n}$ (the 'constrained' model) and one without 
this assumption (the 'unconstrained' model). $N_\phi$, $N_\omega$, $N_\rho$
  indicate the highest radial resonance used in the fit. For the
 resonances above these values their model parameters were used.
 $N_\phi=2$, $N_\rho=3$ for the constrained model and
  $N_\phi=1$, $N_\rho=2$, $N_\omega=2$ for the unconstrained model.
 The unconstrained model fits the data better \cite{Czyz:2010hj},
 but  the quality of the second fit is also reasonable.
 It is interesting to observe that the models give distinct predictions 
  in the region were there are no data available (Fig. \ref{KzKz}),
 even if in the region were data are available their behaviour is
 very similar.

\begin{figure}[ht]
 \vspace{0.5 cm}
\begin{center}
\includegraphics[width=6.5cm,height=6.cm]{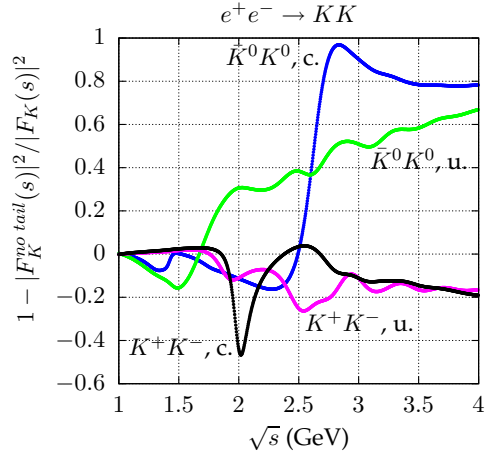}
\caption{ Comparison of the kaon form factors with and
 without ('no tail') infinite tower of resonances. c. denotes the constrained model,
u. the unconstrained one. 
\label{tail}}
\end{center}
 \vspace{0.5 cm}
\end{figure}

It is also of interest to see how much important is the infinite tower
of the resonances in the form factors. In Fig. \ref{tail}  the
relative difference of the form factors without ('no tail') and with the
infinite tower of resonances is shown. In the latter case only $N_i+1 \ , \
i=\phi,\omega,\rho$ resonance contributions are included. As one could
guess, the contributions of the higher resonances to the form factors
are more pronounced for higher invariant masses, but they are not 
negligible even in the lower invariant mass range, where the fitted
 resonances are situated. A similar behaviour
 of the pion form factor was observed in \cite{Czyz:2010hj}.

\section{ Description of narrow resonances }\label{nr}

The narrow resonance (only $J/\psi $, $\psi(2S)$ and $\phi$ were considered)
 contributions were included \cite{Czyz:2010hj} into
 the Monte Carlo generator PHOKHARA for the muon, the charged pion and
 the kaon
 (both charged and neutral) pair production through the following
 substitution 
 $\frac{1}{1-\Delta\alpha(s)} \to C^{VP}_{R,P}(s)\ , \ P
 =\mu,\pi,K^+,K^0\ , \ R = J/\psi,\psi(2S)$ , where 

 \bea
C^{VP}_{R,P}(s) &=& \frac{1}{1-\Delta\alpha(s)}- \frac{3\Gamma_e^\phi}
{\alpha m_\phi} \ BW_\phi(s) \delta_P
 \nonumber \\
 &&+ C_{J/\psi,P}(s)
 + C_{\psi(2S),P}(s)\ ,
\eea

 \bea
C_{R,P}(s) = \frac{3\sqrt{s}}{\alpha}
 \frac{\Gamma_e^R (1+c_P^R)}{s-M_R^2+i \Gamma_R M_R} \ .
\eea

For $P=\mu$ and $P=\pi$, $c_P^R=0$ (it was assumed that
  the narrow
resonances do not decay directly into $\mu^+\mu^-$ and $\pi^+\pi^-$).
 The $\phi$ contributions 
 to the kaon pair production are included in the kaon form factor,
 hence $\delta_K=0$, while $\delta_P=1$ for $P=\mu$ and $P=\pi$.
 The notation and the detailed description of the narrow resonance
 contribution to the amplitude
can be found in \cite{Czyz:2009vj} together with numerical values
 of the $c_P^R$ couplings.

\section{PHOKHARA 7.0}\label{phok}

 The new form factors described in Section \ref{formfactors} and
 the couplings of $\mu^+\mu^-$, $\pi^+\pi^-$, $K^+K^-$ and $\bar K^0K^0$
 pairs to narrow resonances described shortly in Section \ref{nr} were 
 implemented into a Monte Carlo event generator PHOKHARA. It's new
 release (PHOKHARA 7.0) contains also the implementation of the 4-pion
  hadronic currents as described in \cite{Czyz:2008kw}, not included 
 in the previous releases of the code. The code can be found on the web
 page {\tt http://ific.uv.es/\~ \  \hskip - 0.5 cm rodrigo/phokhara/ },
 where also a user guide is provided.

 The variance reduction through standard change of variables using the
  Breit-Wigner shape of resonances was enough to allow for a reasonable
  acceptance rate.
 Other technical problems with the treatment of the narrow resonances
 and their solution were discussed at length in
  \cite{Czyz:2010hj,Czyz:ChPC} and will not be repeated here.
\begin{figure}[ht]
 \vspace{0.5 cm}
\begin{center}
\includegraphics[width=6.5cm,height=5.cm]{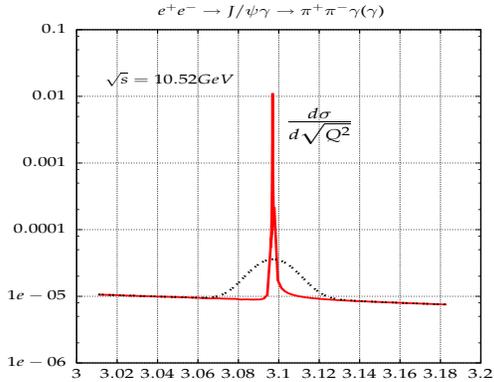}
\caption{ A comparison of the $\pi^+\pi^-$ 'true' (solid
 line)
invariant mass spectra and the $\pi^+\pi^-$ invariant mass spectra seen in
 a detector with a resolution of 14.5 MeV ( as for the BaBar detector 
\cite{Aubert:2003sv}) - dotted line.
\label{resolution}}
\end{center}
\end{figure}
 A typical invariant mass spectrum  of the final meson or muon pair
 in the vicinity of a narrow
 resonance, as seen in a radiative return experiment, resembles more
 a resolution curve of a detector then the shape of the resonance.
  A typical picture is shown in Fig. \ref{resolution}, where the
 generated mass spectrum  of the $\pi^+\pi^-$ pair 
 was convoluted with a Gaussian with a detector
  resolution of the BaBar detector 
\cite{Aubert:2003sv}.
 
\begin{figure}[ht]
 \vspace{0.5 cm}
\begin{center}
\includegraphics[width=6.5cm,height=6.cm]{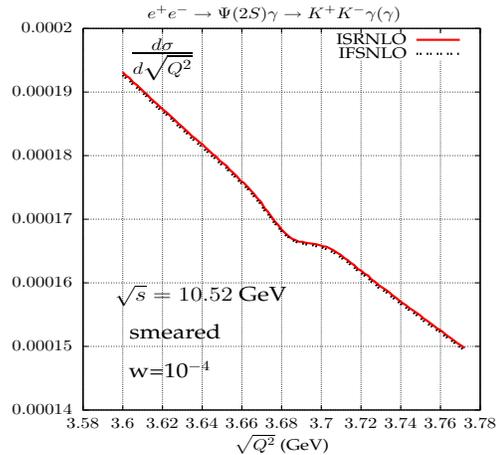}
\caption{  The $K^+K^-$ invariant mass spectra seen in
 a detector with a resolution of 14.5 MeV ( as for BaBar detector 
\cite{Aubert:2003sv}) including only ISR NLO corrections (ISRNLO)
 and adding also FSR NLO corrections (FSRNLO).
\label{psi2s}}
\end{center}
\vspace{-0.5 cm}
\end{figure}

For the $\psi(2S)$ resonance the smearing effects to large extend
dilute the influence of the resonance on the invariant mass spectrum,
 as shown in Fig. \ref{psi2s} for the charged kaon pair invariant mass
distribution. Thus to study it with the radiative return method
 fairly large statistics  is required and it is difficult to expect 
 that the
 method can compete with the accuracy of scan experiments.

\begin{figure} 
\begin{center}
\includegraphics[width=6.5cm,height=6.0cm]{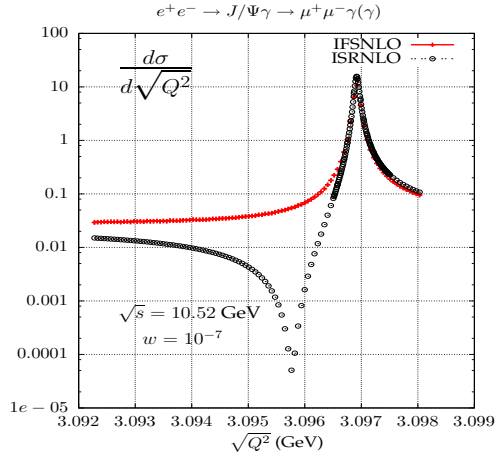}
\caption{ Comparisons of the muon pair invariant mass distributions taking into account only ISRNLO contributions
  and the complete
  (ISR+FSR)NLO result.
}
\label{FSRNLOvsISR}
\end{center}
\end{figure}

\begin{figure} 
\begin{center}
\includegraphics[width=6.5cm,height=6.0cm]{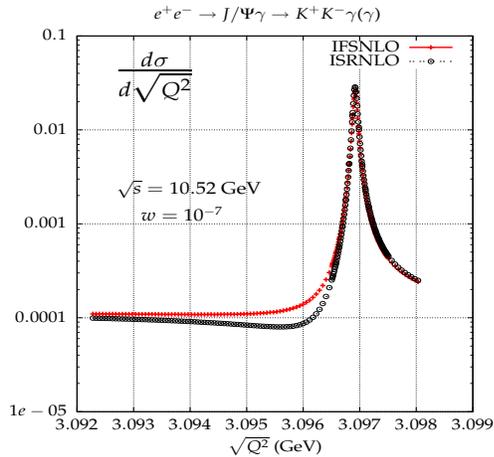}
\caption{ Comparisons of the charged kaon pair invariant mass distributions taking into account only ISRNLO contributions
  and the complete
  (ISR+FSR)NLO result.
}
\label{FSRNLOvsISRkaon}
\end{center}
\end{figure}

\begin{figure} 
\begin{center}
\includegraphics[width=6.5cm,height=6.cm]{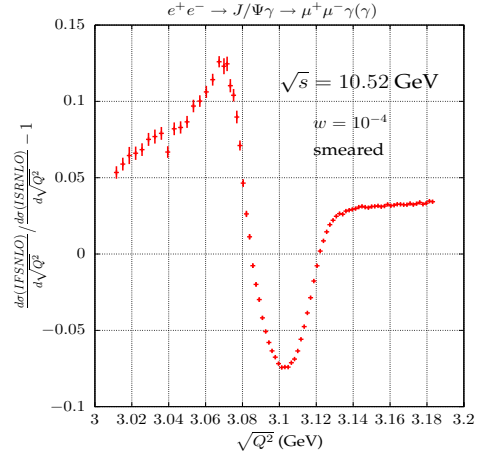}
\caption{ The relative ratio of the muon pair invariant mass distributions 
  taking into account only ISRNLO contributions
  and the complete
  (ISR+FSR)NLO result. Detector smearing effects are taken into account.
}
\vspace{-0.5 cm}
\label{rFSRNLOvsISRxxmu}
\end{center}
\end{figure}

\begin{figure} 
\begin{center}
\includegraphics[width=7.5cm,height=6.5cm]{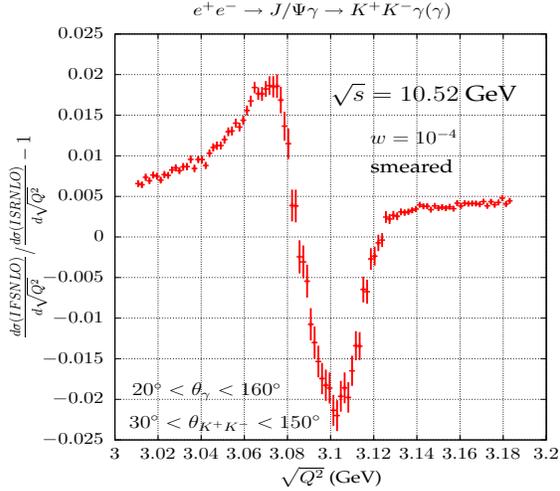}
\caption{ The relative ratio of the charged kaon pair invariant mass distributions 
  taking into account only ISRNLO contributions
  and the complete
  (ISR+FSR)NLO result. Detector smearing effects are taken into account.
}
\label{rFSRNLOvsISRxxka}
\end{center}
\end{figure}

 The contributions from the $J/\psi$ to all studied invariant mass
 spectra is sizable and one can study the $J/\psi$ decays with a good
 accuracy. Thus a legitimate question is how big are the contributions
 coming from the final state photon radiation (FSR). If the detector smearing
 effects are not taken into account, the distortions of the meson
 (muon) pair invariant mass distribution by the FSR effects, as
 compared to ISR, are huge as shown in Figs. \ref{FSRNLOvsISR} and
 \ref{FSRNLOvsISRkaon}. They are still sizable when one takes into 
 account the smearing effects, as evident from
 Figs. \ref{rFSRNLOvsISRxxmu} and \ref{rFSRNLOvsISRxxka}. Moreover they
 do depend on the event selection \cite{Czyz:ChPC}.
However in the integrated cross section to large extend  
 the corrections cancel. For the muon pair production 
$\sigma_{ISRNLO}=6.901$~pb and  $\sigma_{IFSRNLO}= 6.954 $~pb,
while for the charged kaon pair production
 $\sigma_{ISRNLO}=2.450\cdot 10^{-5}$ nb and
  $\sigma_{IFSNLO}= 2.442\cdot 10^{-5}$ nb. Thus when one uses only 
 the integrated spectra  the FSR corrections can be safely neglected.

\section{Summary}\label{sum}
 The implementation into the PHOKHARA Monte Carlo event generator
 of the narrow resonance contributions is presented and the possibility
 of studies of the narrow resonances at B factories is discussed.

\end{document}